\documentclass[final,authoryear,3p,times,twocolumn]{elsarticle}
\usepackage{graphics}
\usepackage{graphicx}
\usepackage{epsfig}
\usepackage{amssymb}
\journal{New Astronomy}
\begin{document}
\begin{frontmatter}
\title{An extensive optical study of V2491 Cyg (Nova Cyg 2008 N.2), from
       maximum brightness to return to quiescence}
\author[um]{U. Munari}
\author[um]{A. Siviero}
\author[sd]{S. Dallaporta}
\author[sd]{G. Cherini}
\author[sd]{P. Valisa}
\author[um]{L. Tomasella}

\address[um]{INAF Astronomical Observatory of Padova, 36012 Asiago
(VI), Italy}
\address[sd]{ANS Collaboration, c/o Osservatorio Astronomico, via
dell'Osservatorio 8, 36012 Asiago (VI), Italy}

\begin{abstract}
The photometric and spectroscopic evolution of the He/N and very fast Nova
Cyg 2008 N2 (V2491 Cyg) is studied in detail.  A primary maximum was reached
at $V$=7.45$\pm$0.05 on April 11.37 ($\pm$0.1) 2008 UT, followed by a smooth
decline characterized by $t^{V}_{2}$=4.8 days, and then a second maximum was
attained at $V$=9.49$\pm$0.03, 14.5 days after the primary one.  This is the
only third nova to have displayed a secondary maximum, after V2362 Cyg and
V1493 Aql.  The development and energetics of the secondary maximum is
studied in detail.  The smooth decline that followed was accurately
monitored until day +144 when the nova was 8.6 mag fainter than maximum
brightness, well into its nebular phase, with its line and continuum
emissivity declining as $t^{~-3}$.  The reddening affecting the nova was
$E_{B-V}$=0.23$\pm$0.01, and the distance of 14 kpc places the nova at a
height above the galactic plane of 1.1 kpc, larger than typical for He/N
novae.  The expansion velocity of the bulk of ejecta was 2000 km/sec, with
complex emission profiles and weak P-Cyg absorptions during the optically
thick phase, and saddle-like profiles during the nebular phase. 
Photo-ionization analysis of the emission line spectrum indicates that the
mass ejected by the outburst was 5.3 10$^{-6}$ M$_\odot$ and the mass
fractions to be X=0.573, Y=0.287, Z=0.140, with those of individual elements
being N=0.074, O=0.049, Ne=0.015.  The metallicity of the accreted material
was [Fe/H]=$-$0.25, in line with ambient value at the nova galacto-centric
distance.  Additional spectroscopic and photometric observations at days
+477 and +831 show the nova returned to the brightness level of the
progenitor and to have resumed the accretion onto the white dwarf.
\end{abstract}


\end{frontmatter}

\section{Introduction}
\label{}

Nova Cyg 2008 N.2 (= V2491 Cyg, hereafter NCyg08-2) was discovered by K. 
Nishiyama and F.  Kabashima at $\sim$7.7~mag on CCD images exposed on Apr
10.73 UT (see Nakano, 2008), and confirmed spectroscopically by Ayani and
Matsumoto (2008).  Immediately following the discovery, it was found that
prior to the outburst NCyg08-2 was an X-ray source (Ibarra and Kuulkers,
2008; Ibarra et al., 2008; Ibarra et al., 2009), detected from the ROSAT
survey era (1990/91) to three months before the outburst (a Swift
observation for Jan 2, 2008).  The only other nova to have been detected in
the X-rays before the outburst was Nova Oph 1998 (=V2487 Oph, Hernanz and
Sala, 2002).  This contributed to trigger a tight X-ray monitoring of the
outburst evolution, which results are described by Page et al., (2008,
2010); Osborne et al., (2008); Ness et al., (2008a,b); Kuulkers et al. 
(2008); Takei et al., (2009); Takei \& Ness (2010).  NOph08-2 displayed
initially a hard X-ray spectrum originating from shocked gas, while a much
brighter and softer X-ray spectrum emerged later.  This super-soft X-ray
emission, that originates from the protracted H-burning during the constant
luminosity phase (Krautter, 2008), ended about 45 days past optical maximum
(Hachisu and Kato, 2009; Page et al., 2010).

IR spectroscopic observations of NCyg08-2 have been briefly described by
Lynch et al., (2008); Rudy et al., (2008); Ashok et al., (2008) and to a
larger extent by Naik et al., (2009).  They found a modest reddening, large
expansion velocities remaining stable over time, slow spectral evolution and
a classification as 'He/N' nova following Williams (1992).

  \begin{figure*}
    \centering   
    \includegraphics[width=9.4cm,angle=270]{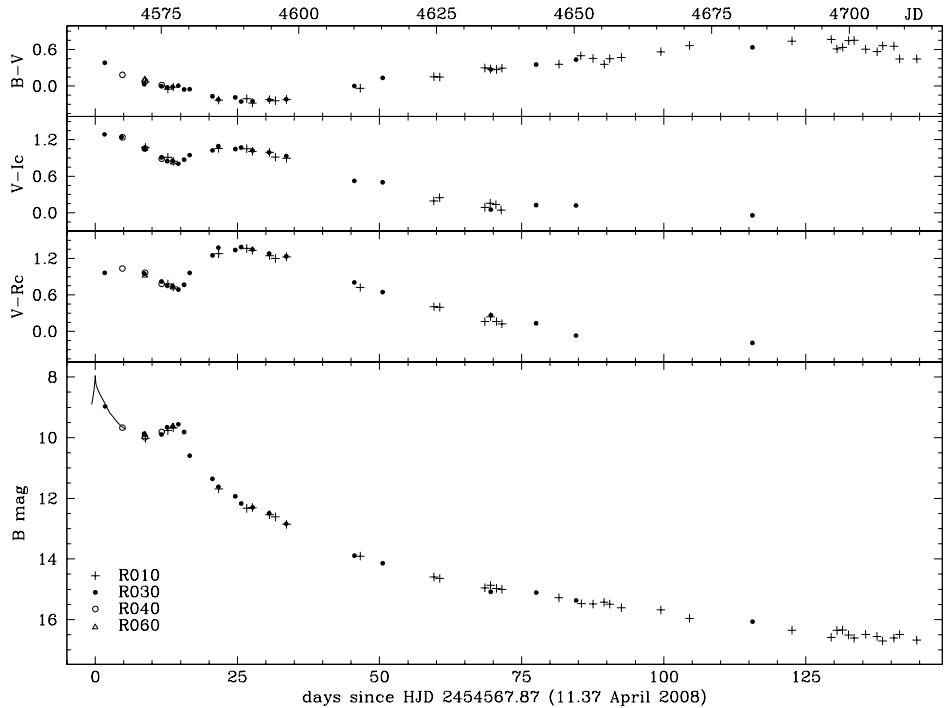}
     \caption{Light and color evolution of Nova Cyg 2008 N.2 
              from our CCD observations. The line around maximum
              brightness in the $B$ band panel is taken from
              Figure~2. Photometry from Table~1, at epochs later than
              day +150, is not included.}
     \label{fig1}
  \end{figure*}  

So far only preliminary descriptions of photometric and spectroscopic
behavior in the optical have been published, and only in the form of 
telegrams/circulars. Tomov et al., (2008a,b) reported about the 
presence, on their low resolution prismatic spectra, of absorption   
components in H$\beta$ and H$\gamma$ at large radial velocities, ranging
from $-$3500 to $-$6400 km/sec depending on the observing date and line.

In this paper we present a detailed study of NCyg08-2 at optical
wavelengths, including a photo-ionization analysis of the ejecta and their
chemical composition, based on our tight photometric and spectroscopic
monitoring of the outburst, that extended from nova discovery well into its
return to quiescence.

\section{Observations}

$B$$V$$R_{\rm C}$$I_{\rm C}$ photometry of NCyg08-2 has been obtained with
several robotic, remotely controlled or manually operated telescopes of the
ANS Collaboration.  Technical details of this network of telescopes and
their operational procedures are presented by Munari et al., (2010a). The
network has been used already for detailed studies of some other recent  
novae (e.g. Munari et al., 2008a,b, 2010b).

All photometric measurements were carefully tied to the local $B$$V$$R_{\rm
C}$$I_{\rm C}$ sequence calibrated by Henden and Munari (2008) against
Landolt's equatorial standards.  The photometry is listed in Table~1 and the
resulting light-curve is presented in Figure~1.  In all, we obtained 66
independent $B$$V$$R_{\rm C}$$I_{\rm C}$ runs distributed over 53 different
nights and spanning an interval of 830 days.  The median value of the
Poisson errors of the photometric points in Figure~1 is 0.005~mag in $V$,
0.008 in $B-V$, 0.006 in $V-R_{\rm C}$, 0.004 in $R_{\rm C}-I_{\rm C}$ and
0.006 in $V-I_{\rm C}$.  The mean r.m.s.  of standard stars from the linear
fit to color equations is 0.019~mag in $V$, 0.029 in $B-V$, 0.027 in
$V-R_{\rm C}$, 0.017 in $R_{\rm C}-I_{\rm C}$ and 0.039 in $V-I_{\rm C}$.

  \begin{table*}
     \caption{Our $B$$V$$R_{\rm C}$$I_{\rm C}$ photometry of Nova Cyg 2008-2.}
     \centering
     \includegraphics[width=11.5cm]{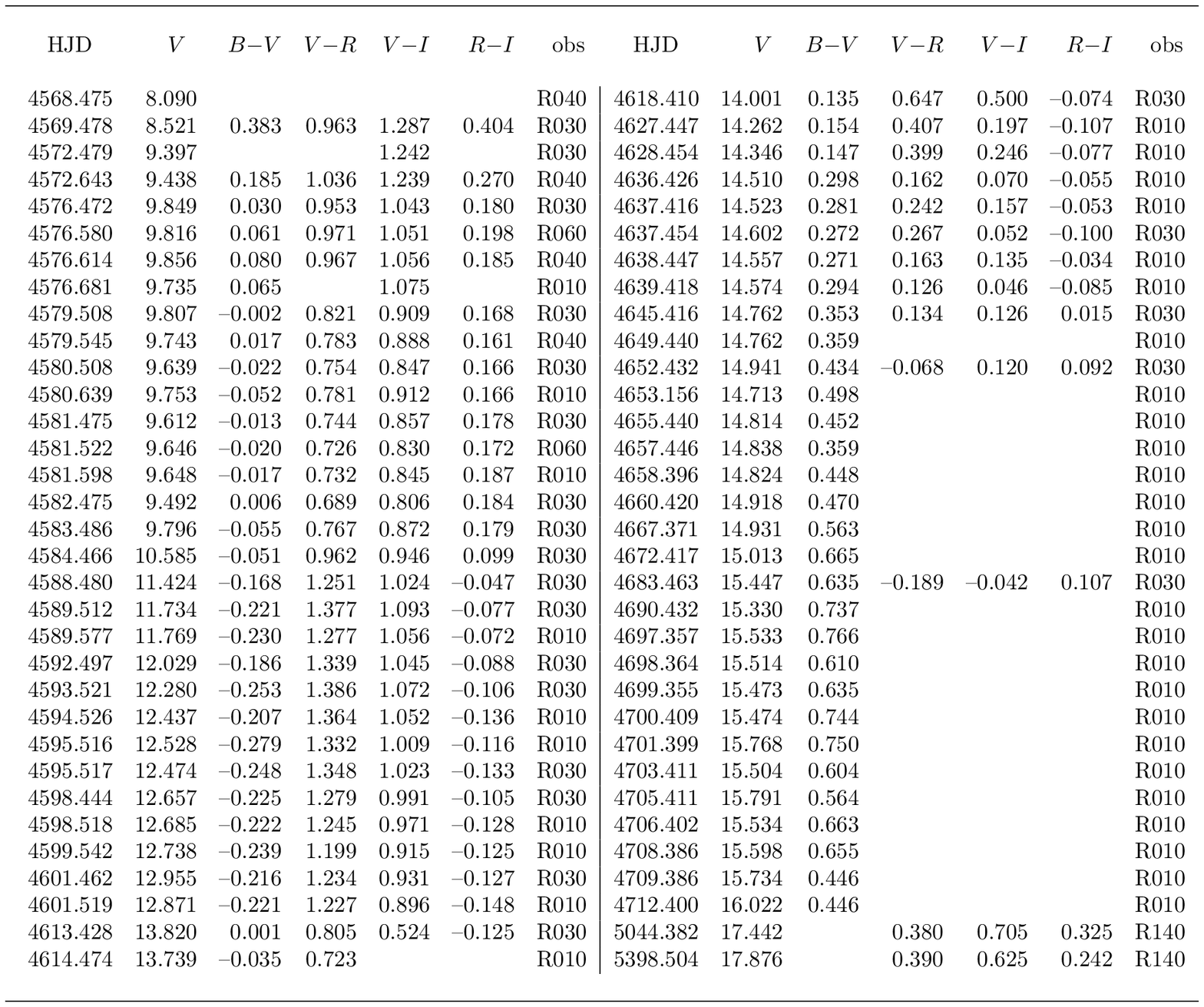}
     \label{tab1}
  \end{table*}

Spectroscopic observations of NCyg08-2 have been obtained with several
telescopes: ($i$) the 3.5m TNG in La Palma (Canary Islands, Spain) and the
high resolution spectrograph SARG, operated at a resolving power of 75\,000,
($ii$) the 1.82m in Asiago equipped with the spectrograph/imager AFOSC with
a 300 ln/mm grism and a 1720 ln/mm volume phase holographic grism, and
($iii$) the 0.6m telescope of the Schiaparelli observatory in Varese
equipped with a multi mode spectrograph and various reflection gratings.  A
detailed journal of the spectroscopic observations is provided in Table~2. 
The spectroscopic data have been reduced and calibrated in IRAF using
standard techniques involving correction for bias, dark and flat fields, and
absolute fluxing using spectrophotometric standard stars observed along
with the nova.  The high accuracy of the absolute fluxes has been checked on
all spectra by integrating the fluxes over the $V$ and $R_{\rm C}$ bands
(whose wavelength ranges are completely covered by our spectra) and
comparing them with photometric data in Table~1.  The differences never
exceeded 0.1 mag for both photometric bands.

\section{Photometric evolution}

\subsection{Rise, maximum brightness and early decline}

The early photometric evolution (first 5 days) of NCyg08-2 is shown in
greater detail in Figure~2.  To draw it, we have used in addition to our
photometry also literature data as indicated.  Some of the literature data
refer to unfiltered CCD observations calibrated against the red USNO-B
magnitudes (rhomb symbols).  They have been transformed into $V$ magnitudes
by applying a rigid shift of +0.68 mag as indicated by the comparison with 
nearly simultaneous true $V$ band data in Figure~2.  This is nicely
confirmed by the photometry of Henden and Munari (2008) for 2005 field stars
around NCyg08-2 that gives for them an average $V$$-$$R_{\rm C}$=+0.61.

The interpolating line in the V panel of Figure~2 has been drawn by hand
to guide the eye, while the lines in the other panels correspond to the
following expressions:
\begin{eqnarray}
B-V       &  = & +0.46 -0.095\times \Delta t + \nonumber \\
            & &   +0.009\times (\Delta t)^2 \\
 V-R_{\rm C}& = & +0.54 +0.310\times \Delta t -  \nonumber \\ 
     & & -0.058\times (\Delta t)^2 +0.003\times (\Delta t)^3 \\
V-I_{\rm C} & = & +1.22 +0.055\times \Delta t -  \nonumber \\ 
        & &   -0.011\times (\Delta t)^2 +0.0001\times (\Delta t)^3
\end{eqnarray}
where $\Delta t$ is the time since maximum in the $V$ band. These behaviors
are normal for novae, for ex.  similar to those displayed by the moderately
slow FeII nova V2615 Oph (N Oph 2007, Munari et al.,  2008a).

The heliocentric  time $t_\circ$ of maximum in the $V$ band is well
constrained in Figure~2 to be April 11.37, 2008 UT, with an uncertainty of
0.1 days.  It will be used in this paper to count the elapsed time.  The
nova reached a maximum brightness of $V$$\sim$7.45 and the decline time was
$t^{V}_{2}$=4.8 days, which corresponds to a classification as {\it very
fast} nova according to Warner (1995).

The rise to maximum has been very fast too, with the last one magnitude
jump completed in 0.6 days (cf Figure~2).  The negative observation by Beize
(2008), who found nothing down to a limiting magnitude of 14 at the position
of the nova on April 8.83, implies that the last 6.5 mag of the rise to
maximum have been covered in less than 2.5 days.

   \begin{table}\footnotesize
      \caption[]{Journal of the spectroscopic observations.}
         \centering 

\begin{tabular}{@{~~}r@{~~~}c@{~~~}c@{~~~}c@{~~}c@{~~}c@{~~~~}c@{~~}}
            \hline
             &       &      &       &         &            \\
        date &  UT   & $\Delta t$     &  expt &       disp  & $\lambda$
range & tel.  \\
             &       & (day)& (sec) &    (\AA/pix)&            \\
             &       &      &       &             &            \\
  2008 04 13 & 01:41 & +1.70&  900  & 1.75         & 3830-7300 & 0.6m \\
       04 13 & 02:32 & +1.74&  900  & 0.30         & 6200-6760 & 0.6m \\
       04 15 & 23:45 & +4.62& 1800  & 1.75         & 3820-7550 & 0.6m \\
       04 16 & 00:57 & +4.67& 3600  & 0.36         & 5465-6240 & 0.6m \\
       04 16 & 02:32 & +4.74&  900  & 1.68         & 6220-8900 & 0.6m \\
       04 22 & 23:14 & +11.6& 1800  & 1.75         & 3880-7550 & 0.6m \\
       04 26 & 05:40 & +14.9&  300  &{\em (75\,000)} & 4620-7920 & 3.5m\\
       05 02 & 22:29 & +21.6& 1800  & 1.75         & 3850-7550 & 0.6m \\ 
       05 02 & 23:40 & +21.6&  900  & 1.68         & 6300-8700 & 0.6m \\ 
       05 14 & 00:32 & +32.7& 1800  & 0.60         & 6175-6835 & 0.6m \\ 
       05 14 & 01:24 & +32.7& 2700  & 3.50         & 3880-7750 & 0.6m \\ 
       07 27 & 21:47 & +108 & 2700  & 4.24         & 3750-7790 & 1.8m \\ 
       07 27 & 22:57 & +108 &  900  & 0.64         & 6400-7050 & 1.8m \\ 
  2009 07 31 & 22:56 & +477 & 7200  & 4.24         & 3700-7770 & 1.8m \\ 
            \hline
   \end{tabular}
   \label{tab2}   
   \end{table}

\subsection{Reddening}

Our high resolution spectrum obtained with the 3.5m TNG telescope on day
+14.9 provides a clean view of the absorption lines of interstellar NaI
toward NCyg08-2.  The profile of the NaI line at 5889.953 is presented in
Figure~3.  It shows several components associated to individual absorption
clouds and/or spiral arms crossed by the line of sight to NCyg08-2.  We have
fitted them with sharp Gaussians, as common practice in high resolution
spectroscopic studies of complex interstellar lines (eg Savage and Sembach, 
1996; Welsh et al., 2010).  The resulting individual Gaussians and the
overall fit are overplotted to the observed spectrum in Figure~3, and their
parameters are listed in Table~3.  Five components are clearly present, with
heliocentric velocities ranging from +4.1 to +49.4 km/s.  Their equivalent
widths have been transformed into the corresponding amounts of reddening
using the calibration by Munari and Zwitter (1997).  The total reddening
affecting NCyg08-2 sums up to $E_{B-V}$=0.24.

van den Bergh and Younger (1987) derived a mean intrinsic color
$(B-V)_\circ$=+0.23 $\pm$0.06 for novae at maximum, and
$(B-V)_\circ$=$-$0.02 $\pm$0.04 for novae at $t_2$.  For NCyg08-2, from
Table~1 and Figure~2, it is $B-V$=+0.46 at maximum and $B-V$=+0.20 at $t_2$,
which correspond respectively to $E_{B-V}$=0.23 and $E_{B-V}$=0.22.

The three estimates, corresponding to observing dates $\Delta t$=0, +4.8 and
+14.9 days, are in perfect agreement, and in the rest of this paper we
will adopt $E_{B-V}$=0.23$\pm$0.01 as the reddening affecting NCyg08-2.

\subsection{Distance}

The rate of decline from maximum and the observed magnitude 15 days past
maximum are calibrated tools to estimate distances to novae.

  \begin{figure}
    \centering  
    \includegraphics[width=8.0cm]{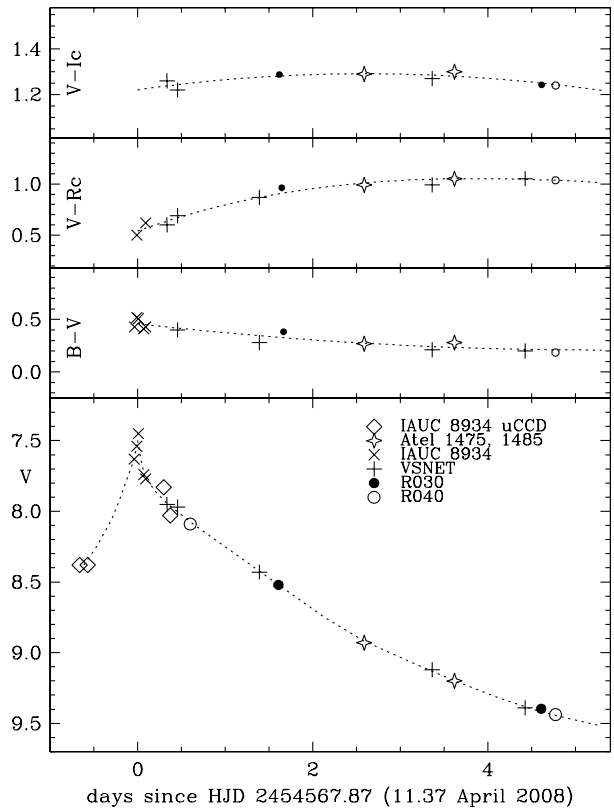}
     \caption{Zooming onto the earliest photometric evolution of Nova Cyg
              2008 N.2 combining our and other data as indicated.}
     \label{fig2}
  \end{figure}   

Published relations between absolute magnitude and rate of decline generally
take the form $M_{\rm max}\,=\,\alpha_n\,\log\, t_n \, + \, \beta_n$.  Cohen
(1988) $M_V$-$t^V_2$ relation provides $M_V$=$-$9.06 for NCyg08-2.  For
$E_{B-V}$=0.23 and a standard $R_V$=3.1 extinction law, this corresponds to
a distance of 14 kpc to NCyg08-2, and to a height above the galactic plane 
of $z$=1.1 kpc.  The shorter distance of 10.5 kpc preliminary derived by   
Helton et al.  (2008), rests on the large $E_{B-V}$=0.43 they adopted from 
Rudy et al.  (2008).  Using instead our more accurate value of
$E_{B-V}$=0.23 would bring Helton et al.  (2008) distance in close agreement
with the 14 kpc we derived.

The $M_B$-$t^B_2$ relations of Capaccioli et al. (1989) and Th.~Schmidt-Kaler 
(cf Duerbeck, 1981) cannot be used with NCyg08-2 because the
re-brightening toward 2$^{nd}$ maximum set in before the nova had declined
by two whole magnitudes in the $B$ band.  For the same reasons, all
relations involving $t_3$ are not applicable to NCyg08-2.

Buscombe and de Vaucouleurs (1955) suggested that all normal novae have the
same absolute magnitude 15 days after maximum light.  However, 15 days after
maximum light corresponds to the time of 2$^{nd}$ maximum for NCyg08-2, and
Buscombe and de Vaucouleurs's relation is therefore not applicable.

  \begin{figure}
    \centering  
    \includegraphics[width=7.8cm]{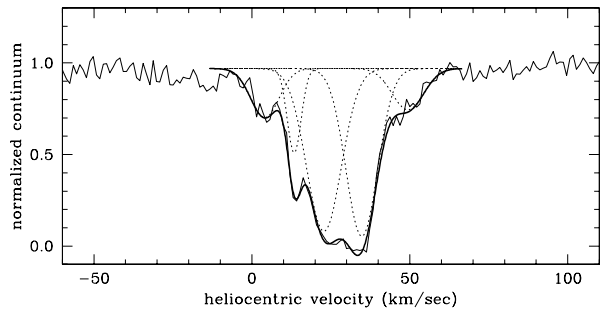}
     \caption{Portion of the SARG spectrum centered on interstellar
              NaI 5889.953~\AA\ line and corrected for telluric absorptions.
              The parameters of plotted Gaussians (dotted lines) are given  
              in Table~3. The thick line is the overall multi-Gaussian fit.}
     \label{fig3} 
  \end{figure}    

  \begin{table}
     \caption{Heliocentric velocity, full width at half maximum,
              equivalent width and corresponding $E_{B-V}$
              for the individual components of the interstellar
              NaI 5889.953~\AA\ line shown in Figure~3.}
     \centering
     \includegraphics[width=4.6cm]{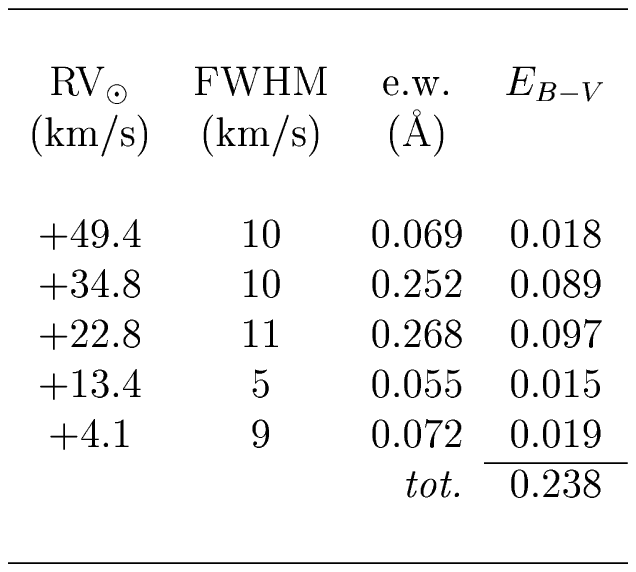}
     \label{tab3}
  \end{table}

The interstellar material causing most of the extinction is concentrated
within 150$-$200 pc of the galactic plane.  The line of sight to NCyg08-2
(at galactic coordinates $l$=67.2, $b$=+4.4 deg) emerges from it at about 2
kpc from the Sun, and it is approximately aligned with the Orion-Cygnus
spiral arm.  According to the Brand and Blitz (1993) maps, the mean
heliocentric radial velocity of the interstellar material along the line of
sight to NCyg08-2 increases up to $\sim$30 km/sec at 2 kpc distance.  This
is the range of velocities observed for the stronger individual components
of the interstellar NaID profile (Table~3).

The galactocentric distance of NCyg08-2 is $R$=13 kpc ($R^2 = R_{\circ}^{2}
+ d^2 - 2 R_\circ d {\rm cos} l$, where $d$=14 kpc is the distance Sun-nova
and $R_{\circ}$=8.5 kpc the galactocentric distance of the Sun).  NCyg08-2 
is therefore located in the external part of the Galaxy, at a significant  
height above the equatorial plane and in a low metallicity ambient.

\subsection{Second maximum and advanced decline}

  \begin{figure}
    \centering  
    \includegraphics[width=8.0cm]{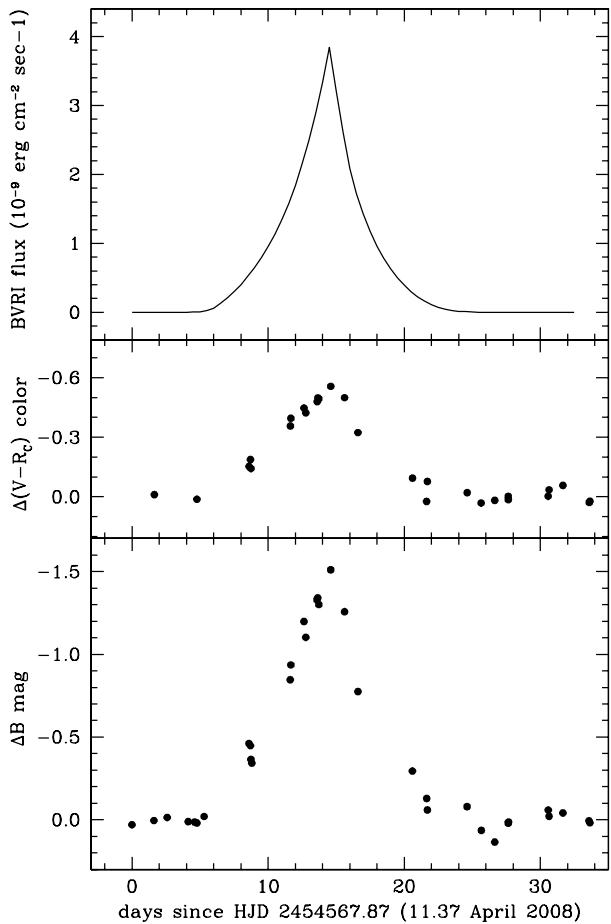}
    \caption{Light-curve, color evolution and flux evolution of the 2$^{nd}$
              maximum.  They are obtained as difference between the observed
              values and the underlying unperturbed decline, extrapolated
              from the photometric behaviour before and after the phase of
              the 2$^{nd}$ maximum.  The flux in the top panel is reddening
              corrected and integrated over the the $B$$V$$R_{\rm C}$$I_{\rm
              C}$ bands.}
    \label{fig4}
  \end{figure}

A relevant feature of the light-curve of NCyg08-2 is the re-brightening it
displayed during early decline, two weeks past the principal maximum.  To
characterize some properties of this 2$^{nd}$ maximum, we have treated it as
the emergence and then the disappearance of an {\em additional source}
(hereafter AS) superimposed onto a normal and smooth underlying decline.
Consequently, we have fitted (with a low degree polynomial) the light-curve
of Figure~1 outside the re-brightening phase and then subtracted it to the
light-curve itself.  The resulting light-curve for AS (i.e.  the photometric
development of the 2$^{nd}$ maximum isolated from the rest) is presented in
Figure~4.

    \begin{figure*}
         \centering
         \includegraphics[width=9.2cm,angle=270]{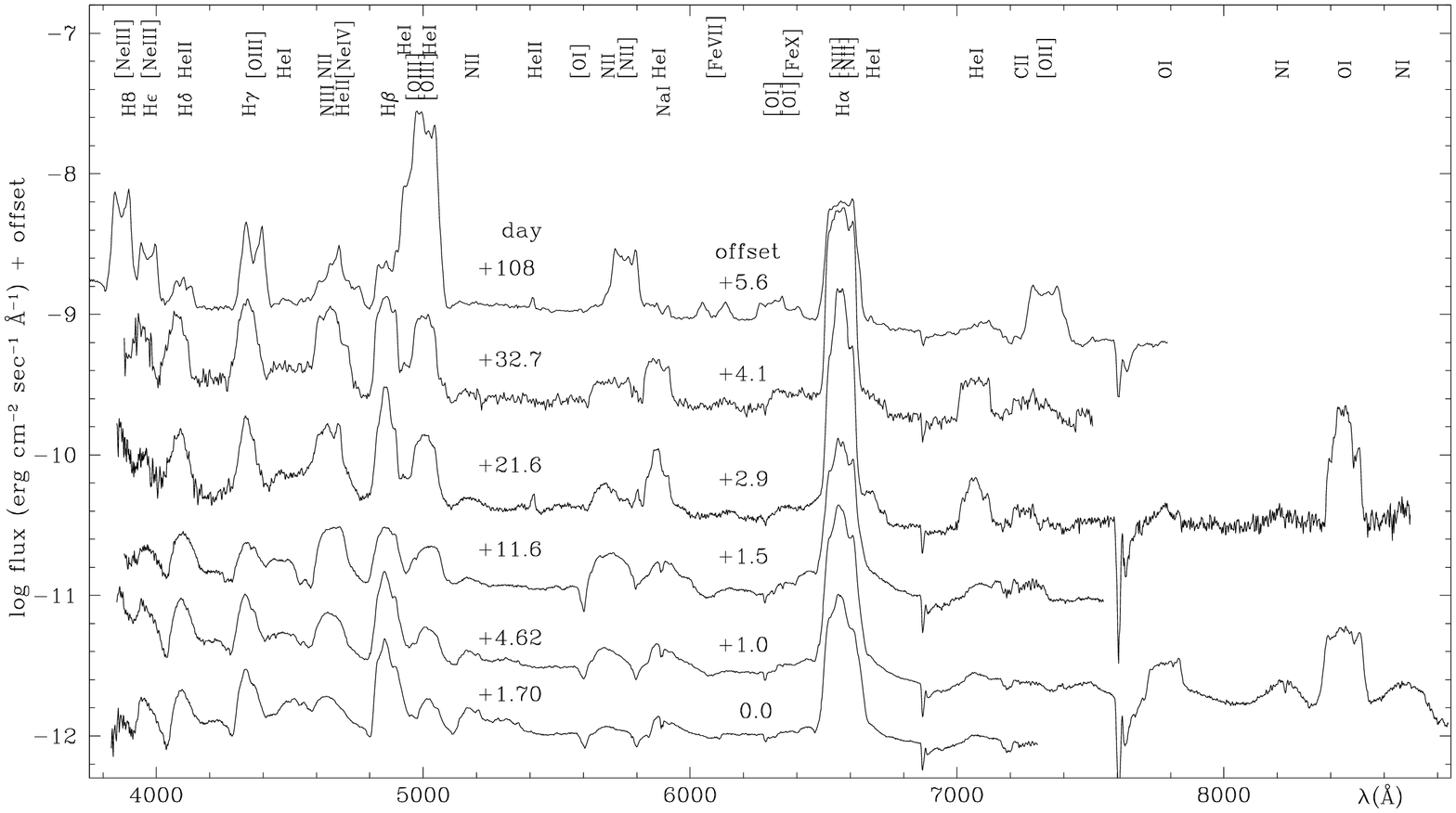}
         \caption{Spectroscopic evolution of Nova Cyg 2008-2. The spectra
         are offset for clarity by the indicated quantity.  The ordinate
         scale is logarithmic to emphasize the visibility of weaker
         features. Figure~10 presents the spectral appearance of Nova Cyg
         2008-2 at later epochs.}
         \label{fig5}
    \end{figure*}

The AS development appears symmetric in the rise and decline branches and of
a shape not much dissimilar from a Gaussian profile.  It started on day
+5.5, well before the nova could have decline by 3 mag from maximum.  The
peak brightness of AS occoured on day +14.5 (April 25.9), and AS ended by
day +24 when the nova set back onto the normal exponential decline from
maximum.  At the time of peak AS brightness, NCyg08-2 became, with respect
to the underlying unperturbed decline, brighter by $\Delta B$=$-$1.5 mag and
bluer by $\Delta (B-V)$=$-$0.02, $\Delta (V-R_{\rm C})$=$-$0.58 and $\Delta
(V-I_{\rm C})$=$-$0.42.  At its peak brightness on day +14.5, the AS
isolated from the rest would have shone at $V$=9.64, $(B-V)$=$-$0.02,
$(V-R_{\rm C})$=$-$0.09, $(V-I_{\rm C})$=+0.18, thus at an absolute
magnitude $M_V$=$-$6.83.  These colors, when corrected for the
$E_{B-V}$=0.23 reddening, are broadly consistent with those of a mid B-type
star.  Assuming that the bolometric correction for AS is the same of a
B5-type super-giant photosphere ({\em B.C.}=$-$1.0 mag, Livingston 2000),
the peak luminosity reached by AS was 1 10$^5$ L$_\odot$, or about 1/3 of
the 3.5 10$^5$ L$_\odot$ of primary maximum.  Integration over time of the
luminosity radiated by AS provides a total of 2.5 10$^{44}$ erg, an amount
equivalent to the hydrogen burning of 2.7 10$^{-8}$ M$_\odot$ of material of
solar composition.

A second maximum has been rarely seen in novae, other two well know cases
are V2362 Nova Cyg 2006 and V1493 Nova Aql 1999a, discussed in detail by 
Munari et al.  (2008b).  The time interval between principal and secondary
maxima for these two novae have been 240 and 45 days, respectively.  That 
seen in NCyg08-2 is therefore occurring much earlier (14.5 days) than in the
other two known cases.

A generally accepted explanation for the secondary maxima is still missing. 
Pejcha (2009) suggested episodic fuel burnings, and Hachisu and Kato (2009)
the release of additional energy associated with rotating magnetic fields. 
However, the lack of a periodic signal in their X-ray observations of
NCyg08-2 argues, for Page et al.  (2010), against the presence of a magnetic
white dwarf in the system.

\section{Spectral evolution}

The spectral evolution of NCyg08-2 is presented in Figure~5. It covers the
period from maximum optical brightness to advanced decline when the nova was
well into the nebular stage (later evolutionary stages are covered in
Figure~10). We also collected high resolution observations of the H$\alpha$
emission line profile at four distinct epochs, and they are presented in
Figure~6. These high resolution H$\alpha$ will not be discussed in detail
in this paper because they are merged into a larger dataset by Ribeiro et
al. (2010) that present a 3D morpho-kinematical model of NCyg08-2 ejecta.

NCyg08-2 displayed strong He and N lines since maximum brightness, with
negligible contribution by FeII lines.  Following Williams (1992), it thus belong 
to the "He/N" class of novae.  These novae tend to be associated with a younger
stellar population, evolve faster, eject less material and harbor more
massive white dwarfs than the novae of the "FeII" type.  The He/N novae lay
closer to the galactic disk than the FeII variant which display an older and
more spheroidal spatial distribution, resembling that of the Bulge (e.g.    
della Valle and Livio, 1998; Shafter, 2008 and references therein).  As a He/N
nova, NCyg08-2 is unusually high above the galactic plane, it laying at
$z$=1.1 kpc, much larger than the scale height of $\leq$100 pc estimated by
della Valle and Livio (1998) for He/N novae.  Other He/N novae high
above the Galactic plane were V477 Sct (= Nova Sct 2005 N2), located at
$z$=0.6 kpc (Munari et al., 2006), or V2672~Oph (= Nova Oph 2009) 
at $z$=0.8 kpc (Munari et al. 2010c).

As typical for very fast novae, NCyg08-2 displayed very broad emission lines and
weak P-Cyg absorption components. P-Cyg profiles were last detected on day
+11.6 on our spectra, and Table~4 summarizes their properties.  From Table~4, the
average FWHM of emission components was 4420 km/sec and the average velocity
shift of the absorption components was $-$4540 km/sec, with however a
significant dispersion among different lines.  This velocity is far larger
than predicted by McLaughlin (1960) relationships for mean velocities of  
both {\it principal} and {\em diffuse enhanced} absorption spectra, which 
predicts $-$1650 and $-$2750 km/sec, respectively.  Tomov et al.  (2008a,b)
reported about a possible P-Cyg absorption component in H$\beta$ at
$-$6400~km/sec on their low resolution, prismatic spectra for day +2.58, and
one at $-$5350~km/sec in H$\gamma$ for day +3.71.  Our larger resolution,   
very high S/N spectra for days +1.7 and +4.7 do not show these high velocity
absorptions, which could have been either spurious or very short lived.
The spectroscopic evolution of NCyg08-2 has been directed toward increasing
excitation conditions along the decline from maximum, as normal for novae. 
On our spectra, HeII 5412 and 4686 \AA\ become visible for the first time on
day +21.6, when the nova was $\Delta B$=3.6 mag down from maximum.  

  \begin{table}
     \caption{Parameters for the absorption lines and the
     corresponding P-Cyg emission components during early evolution
     of NCyg08-2.}
     \centering
     \includegraphics[width=7.8cm]{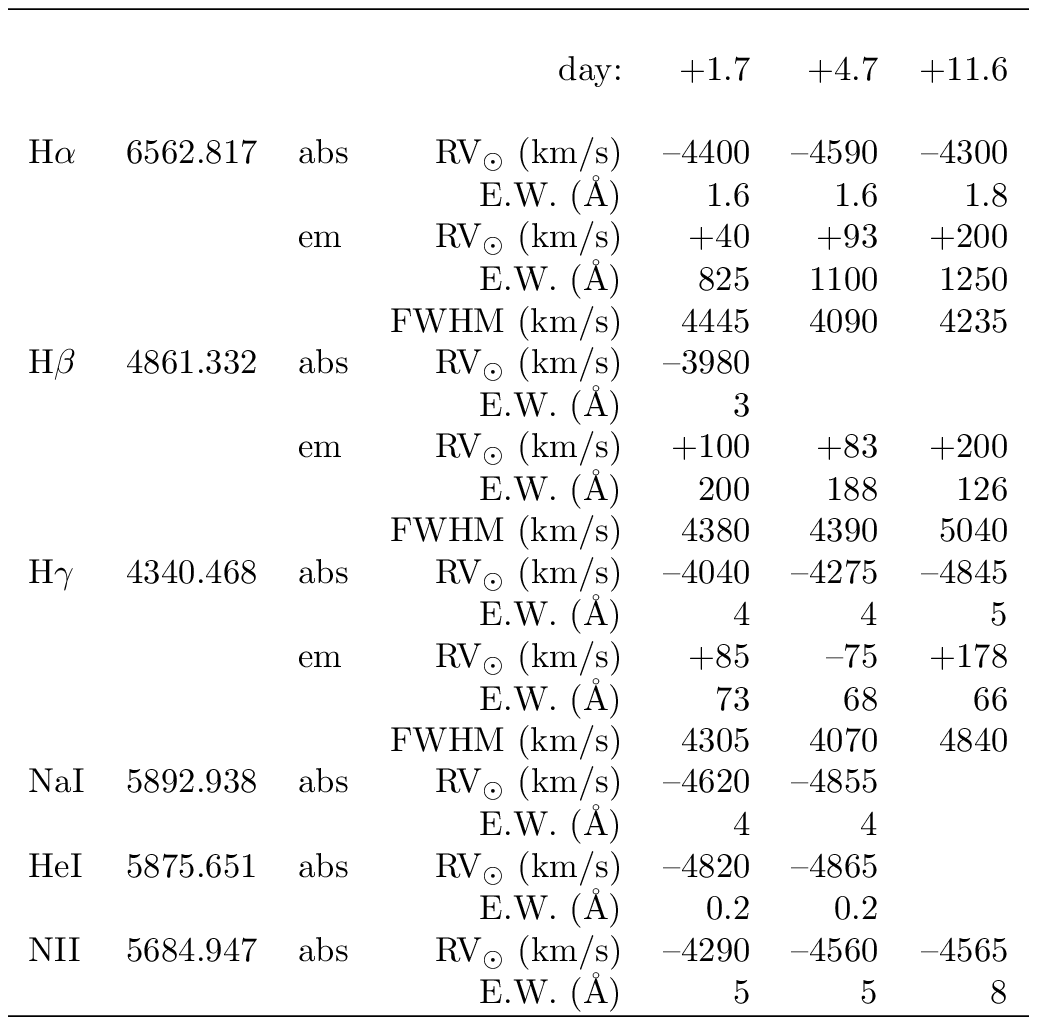}
     \label{tab4}
  \end{table}

  \begin{figure}
     \centering
     \includegraphics[width=7cm]{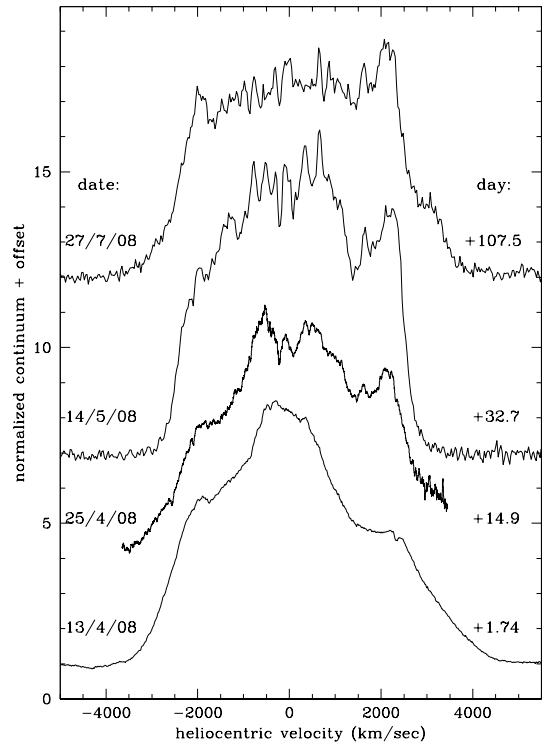}
     \caption{The evolution of the H$\alpha$ of Nova Cyg 2008-2 on our
     high resolution observations (see Table~2).}
     \label{fig6}
  \end{figure}

  \begin{figure}
     \centering 
     \includegraphics[width=6.8cm]{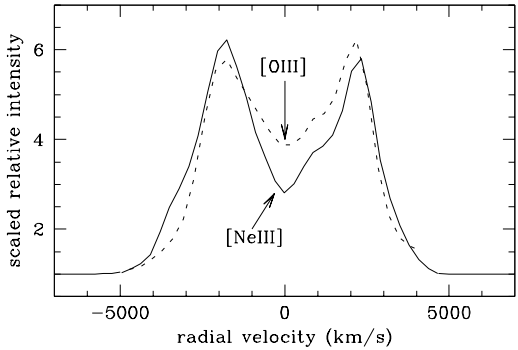}
     \caption{Profiles of [OIII]~4363 and
            [NeIII]~3869 emission lines on the July~27, 2008 (day +108)
            nebular spectrum of Nova Cyg 2008-2.}
     \label{fig7}
  \end{figure}

The large optical thickness of H$\alpha$ during the early outburst phases
and the 2$^{nd}$ maximum is is illustrated by the great intensity of OI
8446~\AA\ in the spectra of Figure~5.  Its intensity under normal
recombination, optically thin conditions should be 0.6 of the OI 7774 line,
which is instead far weaker on day +4.62 spectrum and absent on day +21.6
spectrum.  The inversion in intensity between the two OI lines is usually
associated with fluorescence pumped by absorption of hydrogen Lyman-$\beta$
photons, as first pointed out by Bowen (1947).  For the Lyman-$\beta$
fluorescence to be effective, the optical depth in H$\alpha$ should be
large, presumably owing to the population of the $n=2$ level by trapped
Lyman-$\alpha$ photons.  The $F_{8446}$/$F_{H\alpha}$ flux ration under
optically thin, low ionization conditions and typical nova chemical
abundances is quite low, $\sim$10$^{-3}$ (Strittmatter et al.  1977).  On
days +4.62 and +21.6, the ratio $F_{8446}$/$F_{H\alpha}$ was 0.12
and 0.13, respectively.

On day +108 the nova was well into its nebular stage, and all lines of the
spectrum in Figure~5 for that date were displaying a double peaked,
saddle-like profile typical of an expanding shell or bipolar flow.  Such a
profile for the un-blended [OIII] 4363~\AA\ and [NeIII] 3869~\AA\ lines is
presented in Figure~7, where the velocity separation of the two peaks is
4000 km/sec. It is worth to note that this velocity separation for the two
peaks of the nebular emission line profiles is the same velocity separation
of the shoulders in the H$\alpha$ emission line profile observed close to 
optical maximum brightness (cf spectrum for day +1.74 in Figure~6).

As in most novae, also in NCyg08-2 H$\beta$ evolved unblended with other
major lines during the whole pre-nebular phase.  Figure~8 illustrates the
time dependence of its integrated flux, and how it settled onto the dilution
$t^{-3}$ time scale as soon as the photometric re-brightening, connected to
the 2$^{nd}$ maximum, was over.  The same dilution $t^{-3}$ time scale was
reached, past 2$^{nd}$ maximum, also by the flux through the $V$-band as
illustrated by the lower panel of Figure~8.  Only at later times, the
decline of the flux through the $V$-band started to deviate from the
$t^{-3}$ slope, because of the increasing contribution from the resumed
accretion around the central star (see sect.  7) while the ejecta continued
to fade away. Figure~8 also illustrates the evolution of the
width of H$\beta$, plotted as the width at half of peak intensity (somewhat
different from the FWHM of the Gaussian fitting to the actual profile).  The
evolution of the width has been characterized by two distinct slopes
(plotted as dashed lines in the top panel of Figure 8), with the transition
between the two occurring at the time of 2$^{nd}$ maximum.  During the rise
and decline from 2$^{nd}$ maximum, the width of H$\beta$ (and the other
lines as well) displayed a single expansion/contraction cycle (the solid
line in the top panel of Figure~8), superimposed on the underlying trend.

The highest ionization line in the nebular spectrum of NCyg08-2 on day +108
was [FeX] 6375~\AA.  Its intensity is however too low to qualify that
spectrum as a "coronal" one.  Nevertheless the line is clearly present
with a saddle-like profile. Its expansion velocity is $\sim$1300 
km/sec, lower than that of the other lines, suggesting an origin
in the inner part of the expanding ejecta.

   \begin{figure}
     \centering
     \includegraphics[width=7.5cm]{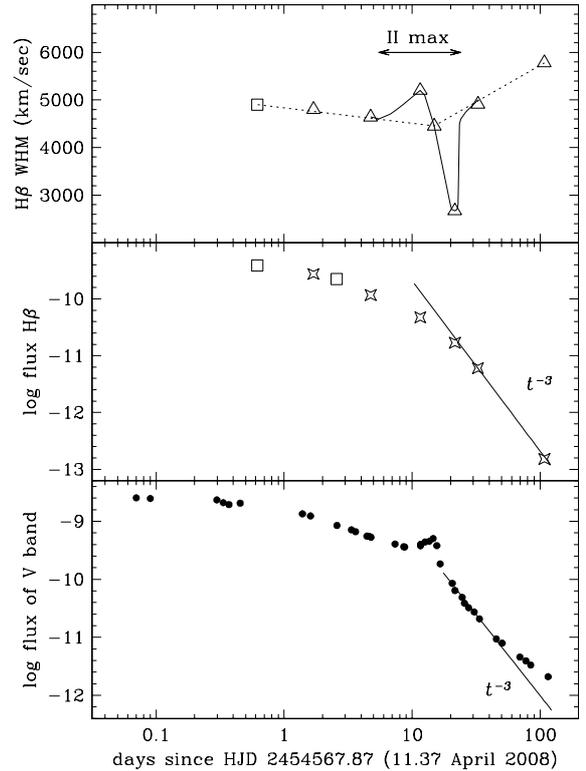}
     \caption{Evolution of the width at half intensity of the
            H$\beta$ emission line (top panel), its integrated
            flux (log of erg cm$^{-2}$ sec$^{-1}$ units; middle panel)
            and integrated flux through the $V$-band (erg cm$^{-2}$
            sec$^{-1}$ units; lower panel).}
     \label{fig8}
  \end{figure}

\section{Photo-ionization analysis}

A photo-ionization analysis of NCyg08-2 nebular spectrum for day +108 has
been performed with the CLOUDY code, version c90 (Ferland et al., 1998),
with the emission line fluxes given in Table~5.  The geometry modeled by
CLOUDY is that of a spherically symmetric shell, with radially variable
density and filling factor.  Given the emerging complexities of nova ejecta,
in particular of the fast novae, with bi-polar structures, equatorial belts,
polar cups and jets, diffuse prolate structures etc.  (see Ribeiro et al.,
2009 for RS Oph,; Woudt et al.  2009 for V445 Puppis; Munari et al., 2010c
for V2672 Nova Oph 2009; Ribeiro et al., 2010 for V2491 Cyg) the spherical
shell geometry adopted by CLOUDY may appear as an over-simplification.  The
parameters derived by CLOUDY should therefore regarded as first order
approximations, useful to frame the overall picture in terms of energetics
of the central star, nebula dimension, chemical composition, mass of the
ejecta.  This is the sense of the photo-ionization analysis carried out in
this section.

  \begin{table}
     \caption{Reddening corrected fluxes of the emission lines
              in the July 27, 2008 (day +108) spectrum of Nova
              Cyg 2008-2. The last two columns express the fluxes
              relative to H$\beta$ as observed and as resulting from
              CLOUDY photoionization modeling (see sect. 5).}
     \centering
     \includegraphics[width=7.2cm]{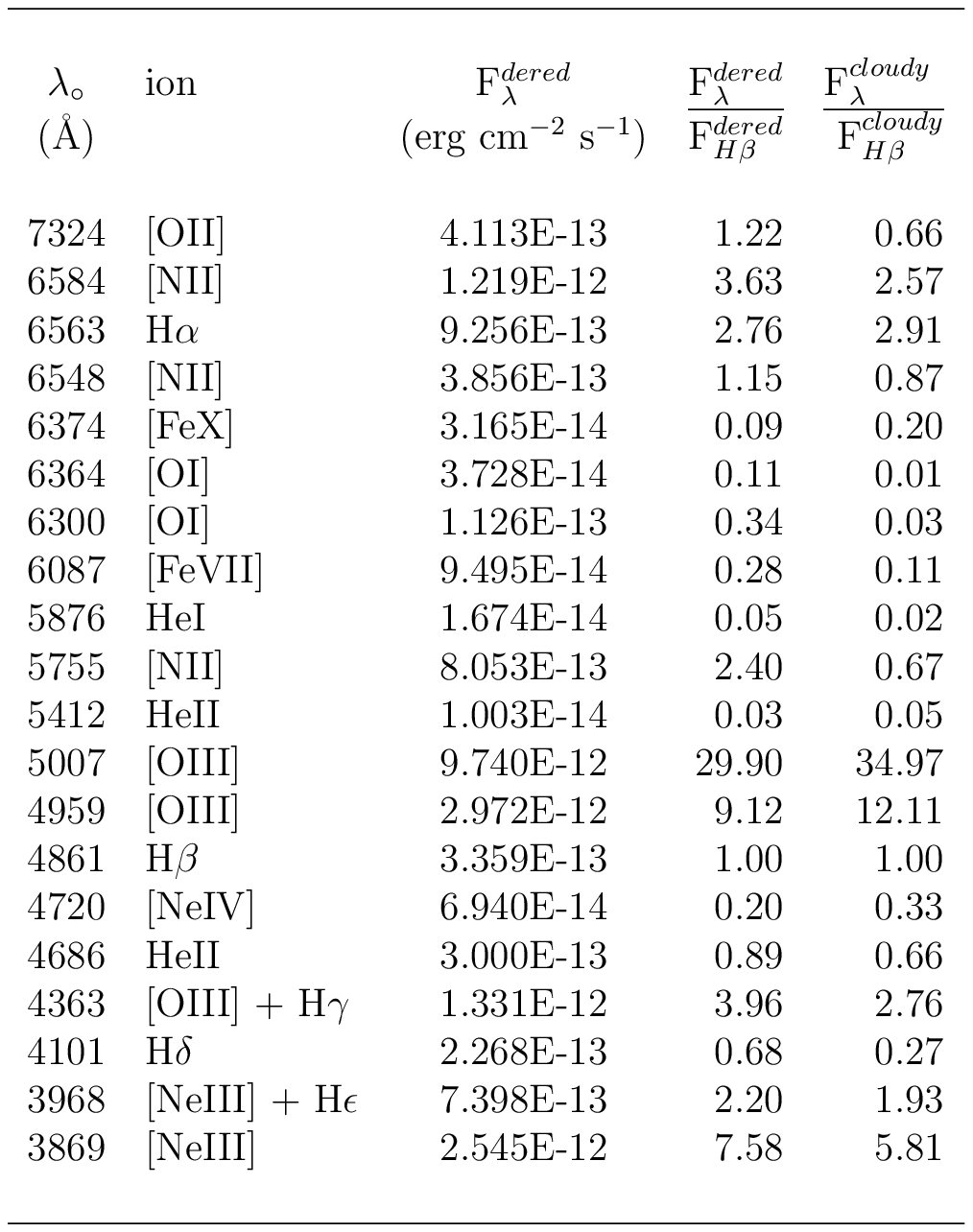}
     \label{tab5}
  \end{table}

We assumed a black body emission for the central star and modeled all lines
in Table~5.  The density profile of a shell, expanding as $r=vt$, is
$\rho(r)$$\propto$$r^{-3}$ and we adopted it.  This is supported by the
results in Figure~8.  We did not fix inner and outer radii for the ionized
shell ($r_{in}$ and $r_{out}$), and treated them as free parameters along
with the covering factor $\omega$=$\Omega$/4$\pi$ (which is the fraction of
the 4$\pi$ sr that is covered by gas, as viewed from the location of the
central star).  Only the abundances of chemical elements with observed lines
were allowed to change, and all others were kept fixed to their solar value. 
The fluxes of the emission lines as resulting from the CLOUDY modeling are
listed in Table~5 together with their observed values. The overall $\chi^2$
of the model is 43.7, or 18.0 if [OIII] 5007 is ignored. Table~6 summarizes 
the modeling results and Table~7 provides the chemical mass fractions.

The shell of ionized gas at day +108 appears to extend from $r_{in}$=85 to
$r_{out}$=180~AU.  Both the inner and outer radii are density boundaries (no
neutral matter external to the shell).  These radii correspond to expansion 
velocities of $\sim$1350 and 2900 km~sec$^{-1}$, respectively.  The velocity
at the inner radius nicely fits that observed for [FeX] (1300
km~sec$^{-1}$), at the outer border the velocity observed for [NII] (2750
km~sec$^{-1}$), and the mean value matches the $\sim$2000 km~sec$^{-1}$  
observed for [OIII] and [NeIII] lines (cf Figure 7).

The central ionizing source is found to have a radius $R$=0.006~R$_\odot$, a
temperature $T_{\rm eff}$=370\,000~K, and therefore a luminosity
650~L$_\odot$, corresponding to $M_{\rm bol}$=$-$2.3.  Both the radius and
the luminosity are smaller than expected during the constant-luminosity
phase of fast novae (Starrfield, 1989; Krautter, 2008), indicating that by day
+108 the stable H-burning at the surface of white dwarf was concluded.  This
is in agreement with evidences from X-ray observations that place the end of
the stable H-burning phase of NCyg08-2 around day +42, according to Page et 
al., 2010, or day +50 following Hachisu and Kato, 2009.

\subsection{Mass in the shell}

The hydrogen mass fraction in the ionized shell is $X$=0.543,
and the covering factor is $\omega$=0.22. Therefore,
the total gas mass within the ionized shell is
  
\begin{eqnarray}
M_{shell} & = & \frac{\omega}{X}\int_{r_{in}}^{r_{out}} 4 \pi r^2 \rho_{\rm H}
(r) dr \nonumber \\
& = & 5.3 \times 10^{-6}  ~~{\rm M}_\odot
\end{eqnarray}

This is very close to the range of ejected mass (from 2.0 to 1.5 10$^{-5}$
M$_\odot$) predicted by various authors (Politano et al., 1995; Starrfield et
al., 1998; Starrfield et al., 2008) for a WD mass of 1.25 M$_\odot$. Hachisu  
and Kato (2009) estimated a mass of 1.3 M$_\odot$ from fitting their models 
to the light-curve of NCyg08-2. This is in agreement with all physical
models
of nova explosion that predict the a mass of the white dwarf closer to
the Chandrasekhar limit with decreasing speed classes $t_2$ and $t_3$ 
(Starrfield, 1989).

The kinetic energy of the ejected shell is

\begin{eqnarray}
E_{kin} & = & \frac{\omega}{X}\int_{r_{in}}^{r_{out}} 2 \pi r^2 \rho_H (r)
\left( \frac {r}{t} \right)^2 dr  \nonumber \\
        & = &  5.6\times 10^{44} {\rm ergs}
\end{eqnarray}
where the velocity of the ejecta is taken to be $v(r) = r/t$ which is
consistent with an ejection over a short period of time around $t$=0.

The outburst had also to provide the energy to unbound the ejected material
from the WD gravitational field.  For a white dwarf of 1.3~M$_\odot$ and
$M_{\rm shell}$=$1.6 \times 10^{-5}$~M$_\odot$ ejecta, the binding energy is

\begin{eqnarray}
E_{bin} & = & G \frac {M_{\rm WD} M_{ej}}{R_{\rm WD}} \nonumber \\ 
        & = & 6.0\times 10^{45}~~~{\rm ergs} 
\end{eqnarray}

The mechanical energy released by the outburst ($E_{\rm bin}$ + $E_{\rm
kin}$) corresponds to the hydrogen burning of 7.0 10$^{-7}$ M$_\odot$ of
accreted matter of solar composition, which is about 13\% of the mass in the
ejected shell.

  \begin{table}
     \caption{Parameters of the CLOUDY model best fitting the de-reddened
              emission line ratio of Table~5.  From top to bottom:
              temperature, radius and luminosity of the central black-body
              source; inner and outer radii of the emitting shell; hydrogen
              density; electronic density and temperature at the inner and 
              outer radii; covering factor; and element abundances relative
              to solar.}
     \centering
     \includegraphics[width=6.5cm]{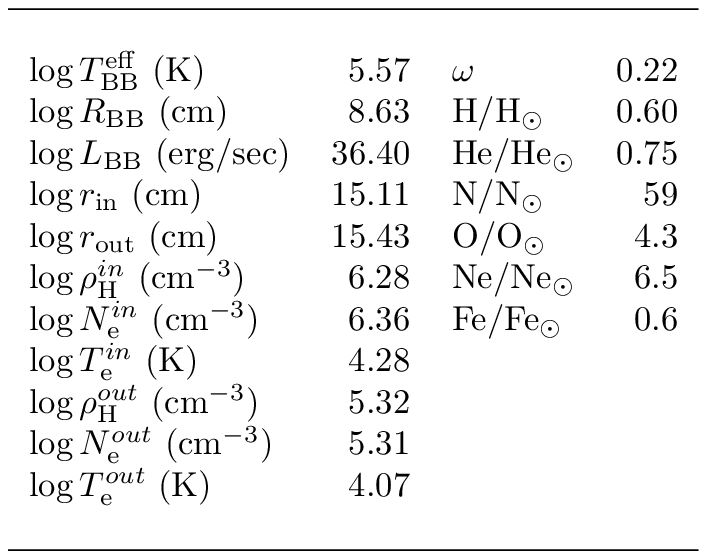}
     \label{tab6}
  \end{table}

\subsection{Chemical abundances}

The chemical mass fractions of NCyg08-2 given in Table~7 reflect the
non-equilibrium CNO-cycle burning of hydrogen (see Gehrz et al., 1998;
Hernanz, 2005), with overabundance of nitrogen and oxygen (no abundance for
carbon was derived because the spectrum at day +108 does not display
measurable lines of carbon ions, and none was expected to be visible given
the prevailing excitation conditions).  The abundance derived for iron,
which is not produced by the TNR, corresponds to a metallicity for the
accreted material of [Fe/H]=$-$0.25.  The sub-solar value agrees with
expected mean ambient metallicity for the Galactic disk
($-$0.5$\leq$[Fe/H]$\leq$$-$0.3, Maciel and Costa, 2010) at the
galactocentric distance (13 kpc) of NCyg08-2, and well within the local
dispersion around the mean value.

There is a clear over-abundance of neon in the ejecta of NCyg08-2. This
element is not produced during the nuclear runaway, but comes from mixing
into the accreted envelope of material from the underlying massive white 
dwarf.  In massive progenitors of white dwarfs, non-degenerate carbon    
ignition leads to the formation of a degenerate core mainly made of oxygen
and neon.  The minimum mass on the zero age main sequence leading to
extensive carbon-burning is $M$$\sim$9.3~M$_\odot$ and the resulting white
dwarf will have a mass of $M_{\rm WD}$$\geq$1.1~M$_\odot$ (e.g. Gil-Pons et
al., 2003). The observed over-abundance of neon thus confirms the evidences 
for a massive white dwarf in NCyg08-2 as inferred by the He/N
classification, the rapid decline and small amount of ejected mass.

Hachisu and Kato (2009) fitted the light-curve of NCyg08-2 with one of their
wind models charecterized by the mass fractions X=0.20, Y=0.48, Z=0.32,
X(CNO)=0.20, X(Ne)=0.10 and X(others)=0.02 for the sum of all remaining
metals.  These mass fractions are not reconciliable with our results in
Table~7.  We tried to fit the observed spectra by adopting the mass
fractions suggested by Hachisu and Kato (2009), but we were not able to 
achieve a satisfactory matching with observations.

  \begin{table}
    \caption{Mass-fraction abundances of measured elements in Nova Cyg 2008
              N.2 and, for reference, in the Sun.}
     \centering
     \includegraphics[width=5.5cm]{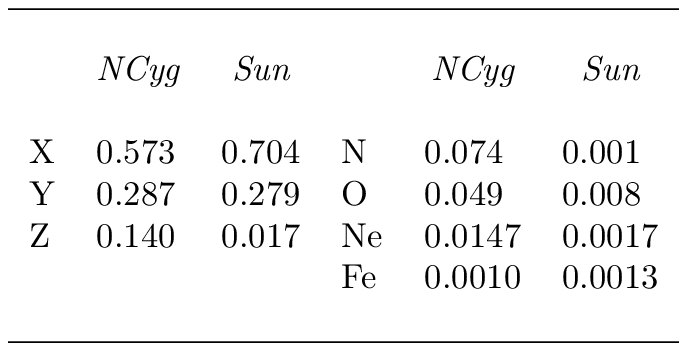}
     \label{tab7}
  \end{table}

\section{The progenitor and the remnant}

We have re-observed NCyg08-2 on 31 July 2009 and 21 July 2010 with the AFOSC
spectrograph+imager mounted on the Asiago 1.82m telescope.  These late
visits at days $t$=+477 and +831 aimed to verify the identification of the
progenitor and the remnant, if the nova had returned to quiescent
brightness, and if accretion had resumed.

The identification and brightness of the progenitor of NCyg08-2 has been
matter of discussion.  IAUC 8934 listed various sources of measurement of
the nova astrometric position, which differ by several arcsec, and reported
conflicting identification with different USNO-B1.0 catalog stars. 
Jurdana-Sepic and Munari (2008) soon after the discovery of the nova
examined historical plates from the Asiago Schmidt telescopes plate archive.
Their measured the only star visible at the position of the nova and
compatible with the available astrometric positions.  They found this star
stable over the period 1970-1986 around mean values $<$$V$$>$=17.06 and   
$B$$-$$V$=+0.82.

A serendipitous monitoring of the field of NCyg08-2 was carried out by
Balman et al.  (2008) from July to November 2007.  They reported that the
monitoring failed to reveal any source at the nova position brighter than
the $R_{\rm C}$=18.2~mag limiting magnitude of their observations.  Balman
et al.  do not specify what is the astrometric position they assumed for the
nova.  They linked their unfiltered CCD observations to the magnitude scale 
to USNO-B1 $R_{\rm C}$ of surrounding stars. By comparing with the Henden   
and Munari (2008) photometric sequence, no systematic offset larger than 0.1
mag is likely to affect USNO-B1 $R_{\rm C}$ values for the region of sky
surrounding NCyg08-2.

  \begin{figure}
    \centering  
    \includegraphics[width=8.0cm]{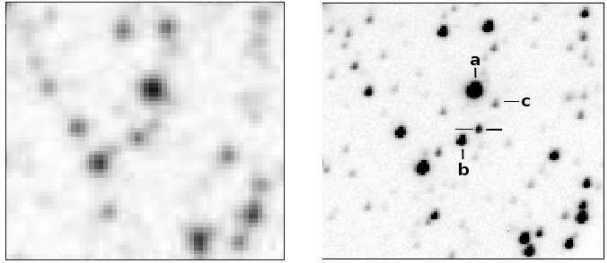}
    \caption{Comparison between the appearance in $I$ band of the nova
             progenitor on SDSS-II (plate exposed on May 25, 1989; left
             panel), and on our CCD observations for 31 July 2009 (day +477),
             at the time of the spectrum in Figure~10, when the nova had
             returned to a brightness equivalent to quiescence.}
    \label{fig9}
  \end{figure}  

    \begin{figure*}
         \centering
         \includegraphics[width=16.0cm]{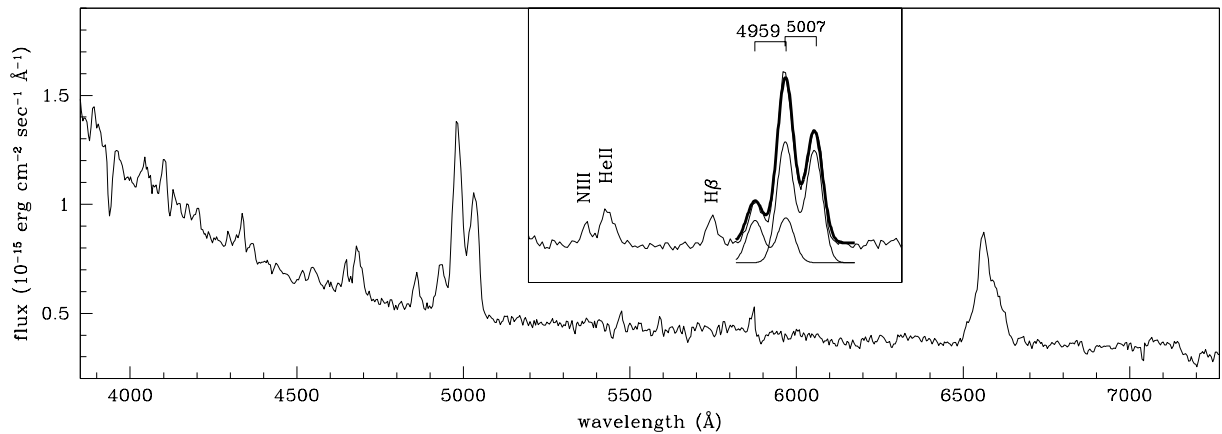}
         \caption{The spectrum of Nova Cyg 2008-2 on 2009 Jul 31 (day +477),
                at the time of the photometric observations of Figure~9.    
                The insert shows the decomposition of the [OIII] 4959,5007
                \AA\ saddle-like profile (see text).} 
         \label{fig10}
    \end{figure*}

Figure~9 compares our AFOSC $I_{\rm C}$-band image from 31 July 2007 with
the corresponding Palomar SDSS-II observation (obtained on 25 May 1989). 
The nova progenitor is barely perceptible on the SDSS-II image.  On the  
AFOSC $I_{\rm C}$-band image the nova and three nearby stars are identified,
for which we have derived the following values: $a$ star $V$=15.40,
$B$$-$$V$=+1.55, $V$$-$$R_{\rm C}$=+0.84, $V$$-$$I_{\rm C}$=+1.50; $b$ star
$V$=17.03, $B$$-$$V$=+0.83, $V$$-$$R_{\rm C}$=+0.58, $V$$-$$I_{\rm
C}$=+1.15, and $c$ star $V$=19.03, $V$$-$$R_{\rm C}$=+0.74 and $V$$-$$I_{\rm
C}$=+1.44.

The values for star $b$ are identical to those found by Jurdana-Sepic and
Munari (2008). This indicates that the star they assumed could be the
progenitor is indeed the field star $b$ in Figure~9. The true progenitor
was too faint to be recorded by the Asiago Schmidt plates (limiting $B$ 
magnitude typically between 18.0 and 18.5). The progenitor is catalogued
as USNO-B1.0~1223-0482965 and it has no 2MASS counterpart. The USNO-B1.0
magnitudes appear unreliable for the stars in the immediate vicinity and
for the progenitor itself, most probably an effect of the crowding in the
field. For ex., contrary to evidence from direct inspection of the SDSS  
plates and results of CCD observations, the USNO-B1.0 catalogue gives    
the same $B$$\sim$16.2 mag for both $a$ and $b$ stars, instead of        
respectively $B$=16.95 and $B$=17.86.

We have then estimated directly on Palomar SDSS-II images the brightness of
the nova progenitor and found: $B$$\sim$18.3, $R_{\rm C}$$\sim$17.4, $I_{\rm
C}$$\sim$16.9 (uncertainties $\pm$0.2 mag).  On AFOSC images for 31 July
2009 the nova shined at $V$=17.44, $R_{\rm C}$=17.06 and $I_{\rm C}$=16.73,
and $V$=17.88, $R_{\rm C}$=17.49 and $I_{\rm C}$=17.14 on 21 July 2010
(uncertainties $\pm$0.03 mag).  Thus, at the time of the photometric and
spectroscopic observations on days +477 and +831 the nova had returned to a
brightness close to that of quiescence.  Finally, it has to be noted that
Balman et al.  (2008) report that the progenitor was fainter than 18.2 in
$R_{\rm C}$ band in 2007 is equivalent to say that it was fainter than star
$c$ in Figure~10.  This was not the case at the time of the Palomar SDSS-II 
$R_{\rm C}$ image in Figure~9.

\section{Resuming the accretion}

The spectrum of NCyg08-2 on day +477, at a time when the nova had already
returned close to quiescence brightness, is presented in Figure~10. It is
characterized by a hot continuum and high excitation emission lines,
with the intensity of HeII 4686 \AA\ slightly larger than that of H$\beta$.

Two sets of lines are simultaneously present on the day +477 spectrum: (1)
nebular lines from highly diluted, distant and expanding material,
and (2) permitted lines from resumed accretion. 

The first type of lines is exemplified by the saddle-like profiles of [OIII]
4959, 5007 \AA\ lines.  The insert of Figure~10 illustrates the
de-convolution of the [OIII] 4959, 5007 blend with two individual
saddle-like profiles characterized by a velocity separation of 3050 km/sec
of the two peaks.  A similar de-convolution with a similar velocity
separation works well for the [NII] 6548, 6584 \AA\ blend, with in addition
a single peaked H$\alpha$ component superimposed.  At the time of the
nebular spectrum for day +108 in Figure~5, the velocity separation of the
two peaks of the saddle-like profile of [OIII] lines was 4000 km/sec.  Thus,
in the intervening year, ($i$) the expansion of the ejecta has been either
slowed down or ($ii$) the emission from [OIII], [NII] lines at day +477 came
from more internal and therefore slower layers than [OIII], [NeIII] at day
+108 in Figure~7.  The first possibility requires a large deceleration of
the ejecta and is not plausible.  Supposing that ($a$) this occoured at a
uniform rate in the time interval between day +108 and +477, and that ($b$)
it involved the whole ejected shell, then the energy radiated by the
associated shock front would have been $\sim$1.7 10$^{44}$ erg for a mean
luminosity of 5.5 10$^{36}$ erg/sec $\equiv$ 1400 L$_\odot$.  Considering
that the temperatures associated to shock fronts bring the peak of the
emitted energy into the most energetic part of the electromagnetic spectrum,
NCyg08-2 should have been a very bright and a very hard X-ray source during
the time interval between day +108 and +477, contrary to evidence from the
Swift observations by Page et al.  (2010) that extend to day +236.  The
second possibility is instead in line with the fact that the emissivity of
lines depend from electron density, which declines as $r^{-3}$, thus faster
in the outer ejecta that expand at higher velocities.

The second type of lines visible on the day +477 spectrum of Figure~10,
which are characterized by a single-peaked and sharp profile, is composed by
hydrogen Balmer series, HeI, HeII and NIII.  These lines and the hot
underlying continuum closely resemble those of cataclysmic variables, close
matches being for example the spectra of BO Cet (Zwitter and Munari, 1995) or
that of the old novae RR Pic (Williams and Ferguson, 1983) and HR Del (Munari
et al.,  1997).  This close similarity with CV spectra supports the idea that
accretion had already resumed at day +477 on NCyg08-2.  The short time scale
flickering of X-ray emission detected by Page et al.  (2010) in NCyg08-2 at 
advanced evolutionary phase, when the hydrogen burning and super-soft phase 
was already over, is also supporting the fact that NCyg08-2 had resumed
accretion at the time of our day +477 spectroscopic observation.  

\section{Acknowledgments}

We would like to thank Andrea Frigo, Paolo Ochner, Flavio Castellani,
Stefano Tomasoni and Valeria Luppi of the ANS Collaboration for their   
assistance in the acquisition and treatment of part of the data 
presented in this paper.

\end{document}